\documentclass[usenatbib]{mn2e}
\usepackage{times}
\usepackage{epsfig}
\usepackage{amsmath}
\usepackage{amssymb}
\usepackage{longtable}
\title[The strong environmental dependence of black hole scaling relations]{The strong environmental dependence of black hole scaling relations}

\author[S. L. McGee]{Sean L. McGee\thanks{Email: mcgee@strw.leidenuniv.nl} 
 \\
Leiden Observatory, Leiden University, PO Box 9513, 2300 RA 
Leiden, The Netherlands\\ 
}

\date{\today}

\def\LCDM{$\Lambda$CDM$~$}

\def\Mdot{M$_\odot$}
\def\Msun{M$_\odot$}
\def\Lsun{L$_\odot$}
\def\Ks{K$_{\rm S}$}
\def\veldisp{$\sigma$}
\def\Mbh{M$_\bullet$}
\def\Mbulge{M$_{\mathrm{bulge}}$}
\def\Mlum{M$_{\mathrm{V,bulge}}$}
\def\mbhbullum{M$_\bullet$ - M$_{\mathrm{V,bulge}}$}

\def\Mbhrel{M$_\bullet$ - $\sigma$}
\def\Mbhbulge{M$_\bullet\ $ - $M_{bulge}$}
\def\Mpc{Mpc}

\def\kms{${\rm km}{\rm s}^{-1}$}
\def\kmsmpc{\>{\rm km}\,{\rm s}^{-1}\,{\rm Mpc}^{-1}}

\newcommand{\degree}{\ensuremath{^\circ}}

\begin{document}

\maketitle
 
\begin{abstract}
We investigate how the scaling relations between central black hole mass and host galaxy properties (velocity dispersion, bulge stellar mass and bulge luminosity) depend on the large scale environment. For each of a sample of 69 galaxies with dynamical black hole measurements we compile four environmental measures (nearest neighbor distance, fixed aperture number density, total halo mass, and central/satellite).  We find that central and satellite galaxies follow distinctly separate scalings in each of the three relations we have examined. The \Mbhrel\ relation of central galaxies is significantly steeper ($\beta$ = 6.38 $\pm$ 0.49) than that of satellite galaxies ($\beta$ = 4.91 $\pm$ 0.49), but has a similar intercept. This behavior remains even after restricting to a sample of only early type galaxies or after removing the 8 brightest cluster galaxies. The \Mbhrel\ relation shows more modest differences when splitting the sample based on the other environmental indicators, suggesting that they are driven by the underlying satellite/central fractions. Separate relations for centrals and satellites are also seen in the power law scaling between black hole mass and bulge stellar mass or bulge luminosity. We suggest that gas rich, low mass galaxies undergo a period of rapid black hole growth in the process of becoming satellites. If central galaxies on the current \Mbhrel\ relation are representative progenitors of the satellite population, the observations imply that a $\sigma$ = 120 \kms\ galaxy must nearly triple its central black hole mass. The elevated black hole masses of massive central galaxies are then a natural consequence of the accretion of satellites. 

\end{abstract}


\section{Introduction}

The correlation between the mass of a central black hole (BH) and the properties of its host galaxy suggests black holes play a major role in galaxy evolution and has stimulated a wide range of theoretical and observational studies. It is now clear that the mass of the black hole forms a tight relation with the velocity dispersion of its galaxy \citep{Ferrarese00, Gebhardt00, tremaine_bhm, Graham2011, McConnellMa2013}, the mass of its bulge \citep{Magorrian97, marconi_hunt, haring_rix}, the luminosity of its bulge \citep{KormendyRichstone, KormendyGebhardt, Graham2007, GrahamScott2013} and the Sersic profile of the bulge \citep{Graham2001, GrahamDriver}. It has been postulated that these tight relations point to a self-regulation between the growth of the galaxy and its black hole. Feedback from an active galactic nucleus (AGN) is a plausible mechanism for this self-regulation and may explain the quenching of star formation in massive galaxies \citep{silk_rees, kauffmann_haehnelt, King03, springel_agn, croton, bowermodel, hopkins_agn1, hopkins_agn, BoothSchaye}. In addition, such feedback may explain the slope of x-ray scaling relations in groups and clusters as well as the large cavities observed in their x-ray surface brightness distribution \citep{mcnamara, Bowerxray, MccarthyAGN, dubois_2012}. Nonetheless, many questions remain about how the AGN manages to accrete gas \citep{CattaneoTeyssier, hopkins_gas, pizzolato_soker, dubois_feeding} and how the AGN outflow couples to the surrounding gas \citep{Murray05, ciotti_ostriker, ostriker10, novak12, fauchergiguere, silk_nusser}.

A powerful way to test models of black hole growth is by examining how their scaling relations change in different environments. It has been known for some time that galaxies in dense environments, tend to be passive, spheroidal and lacking in gas \citep{dressler80, HGC_84, postman, balogh_cnoc1, blanton05, poggianti, wilmanS0, Bamford_zoo, Peng10, mcgee_11, wetzel}. It is now clear that galaxies being accreted into dense environments must halt their star formation and alter their morphology. Although the exact physical mechanism responsible is not known, it may be directly linked to the hierarchical formation of galaxies. It is thought that galaxies form at the centre of dark matter halos, after the baryonic material has cooled, condensed and fragmented into stars \citep{rees_ostriker, white_rees, Blumenthal1984}. The continual feeding of gas and smaller galaxies into this central location results in continued formation of stars and/or growth of the galaxy stellar mass in these galaxies, often called central galaxies. In contrast, the hierarchical growth of structure in the Universe results in dark matter halos with galaxies in their centres being accreted into larger halos, where they spend some time orbiting the potential \citep{press_schechter, lacey_cole}. The details of the baryonic processes important in the further evolution of these `satellite' galaxies is murky, but it seems likely that a satellite undergoes a stripping of its own gas as it encounters the hot medium within groups and clusters \citep{gunn, adabi, Quilis_S0, mccarthy, bahe}. 

The morphological transformation associated with being accreted into a dense environment is especially important for the black hole scaling relations because it may provide a new channel to the formation of bulges and early type galaxies. Unfortunately, how galaxies transform from spirals to early types is a matter of some disagreement in the literature. It has been suggested that the bulges of early types are too large to come from spiral galaxies without some additional bulge growth \citep{christlein04, Bekki_bulge, lackner_gunn}. However, large samples of resolved color profiles are compatible with a gentle fading of disks and the revealing of their prexisting bulges \citep{weinmann_faded, hudson_faded}. Either way, this unique route of galaxy evolution is directly related to dense enviroments and thus is a powerful testbed for the nature of black hole scaling relations. 

While the observational definition of `satellite' galaxies varies, the motivation behind it is to separate `accreted' galaxies from the main galaxy of the halo. Individual galaxy groups often have poorly constrained centers and thus the identification of a galaxy as a `central' or `satellite' is made based on the relative stellar mass or luminosity of the galaxies and not their specific location \citep{YMvJ, weinmann,George2011}. This alternative definition is partially motivated by the apparently tight relation between central stellar mass and dark matter halo mass \citep{Mandelbaum2006, more2009, moster_hod, Behroozi2010}. If the relation between central stellar mass and dark matter halo mass was monotonic with zero scatter then central galaxy would always have the most stellar mass. Despite the possible confusion, it has become common in the literature to refer to galaxies as `central' or `satellite' even if the selection does not explicitly depend on location \citep[eg.][]{weinmann, wetzel}. Nonetheless, simply designating the galaxy with the most stellar mass in a group as the central results in radial and velocity positions consistent with those galaxies being at rest at the centre of the group halo \citep{vandenbosch_phase, skibba_phase}.

The distinction between a central galaxy and satellite galaxy may be most important for understanding galaxy evolution, however, there are reasons to believe as many as three types of `environment' other than central/satellite might be important for black hole scaling relations. Firstly, the processes of harassment and merging rely largely on interactions with other galaxies, which implies that the likelihood for interaction, or the distance to an Nth nearest galaxy, is important. Secondly, there is evidence that dark matter halos form sooner in large scale overdensities on the order of $\sim$ 5-10 \Mpc\, so that the earliest forming 10$\%$ of halos at fixed mass are 5 times more strongly clustered than the latest forming 10$\%$ \citep{limogao, gaospringelwhite}. If this growth is directly linked to the formation of galaxies, then it is possible that galaxies in large scale overdensities have a `head-start' in their development. Finally, the total mass of a galaxy's host halo, regardless of the galaxy's location, may influence how the gas is accreted onto the galaxy \citep{simha, zavala12}. 

In this paper, we calculate several environmental measurements for each galaxy with a direct black hole measurement as compiled by \citet{McConnellMa2013}. Each of these environmental measurements is designed to probe a potentially important environment. We will use these measurements to divide the sample of black holes into subgroups and examine the resulting scaling relations. In \textsection \ref{sec-data} we discuss the black hole and galaxy property measurements we use and how we derive the environmental measures. In \textsection \ref{sec-results} we show the strong effect environment has on the black hole scaling relations and in \textsection \ref{sec-discuss} we discuss how this could arise. We conclude in \textsection \ref{sec-concl}. In this paper, we adopt a \LCDM cosmology with the parameters; $\Omega_{\rm m} = 0.3$, $\Omega_{\Lambda}=0.7$ and $h=H_0/(100 \kmsmpc)=0.70$. 

\section{Data and Methods}\label{sec-data}
We plan to examine the scaling relations between black holes and their internal galaxy properties with the aim of determining the effect of their external, or large scale, environment on such relations. The large scale environment profoundly affects many galaxy properties, such as star formation rate and morphology. Indeed, it is this effect which makes examining the black hole scaling relations in different environments a worthwhile endeavour. However, for this reason we must restrict our sample of black hole masses to those which are determined as directly as possible, that is, those derived from stellar or gas dynamics. Several indirect measurements, such as those based on the width of emission lines, exist \citep{mclure_dunlop, shen_bias}, but these are calibrated on the BH scaling relations and could also be systematically affected by environmental conditions. Thus, we will restrict our discussion to low redshift scaling relations where the black hole mass has been more directly measured.

\subsection{Black holes and their host galaxy properties}
We take as our starting point the recent compilation of \citet{McConnellMa2013}. These authors have presented critical evaluations of nearby black hole and host galaxy properties accompanied by judicial pruning based on the quality of the measurements. As a base, we will use the black hole mass, velocity dispersion, $V$ band bulge luminosity and bulge stellar mass of their full sample of 72 galaxies. While each galaxy has a measurement of the black hole mass and velocity dispersion, only 35 of the early type galaxies have bulge stellar mass measurements and 44 have bulge luminosities. The data is derived from a large literature, and as such the detailed method for each measurement varies. The black hole masses are derived from either stellar dynamics (48), gas dynamics (14) or maser dynamics (10). The bulge $V$-band luminosities were reported for ellipticals and S0 galaxies for which the bulge and disk could be cleanly separated. The bulge stellar mass was determined by multiplying the bulge $V$-band luminosity by the dynamically determined $M/L$. In some cases, the $M/L$ had to be converted to $V$-band using observed galaxy colours before determining the stellar mass. In addition to this uncertainty, many of the bulge masses were derived by assuming that mass follows light, which may explain inconsistencies in M/L measured from dynamics and stellar populations \citep{cappellari2006, conroy2012}. For these reasons, McConnell $\&$ Ma have assumed an error of 0.24 dex in the measurement of \Mbulge.

Of this base sample of 72, as discussed in \textsection \ref{sec-enviro}, we were not able to make environmental measures for two galaxies in the sample (the Milky Way and Circinus). Another galaxy, M32, is of significantly lower velocity dispersion, luminosity and mass than the rest of the sample and could bias the scaling relations when the sample is split in two. Therefore, we remove all three galaxies from further analysis, leaving a reduced sample of 69 galaxies. The only further adjustment we make to the tabulated values presented in McConnell $\&$ Ma is in the black hole mass of NGC1399. For this galaxy, the authors have presented two measurements for the black hole mass, however, we have chosen to average these values to \Mbh = 9.1$^{+2.8}_{-3.9}$$\times$10$^{8}$ \Mdot.

\subsection{Environmental measures}\label{sec-enviro}

\begin{figure}
\leavevmode \epsfysize=5.6cm \epsfbox{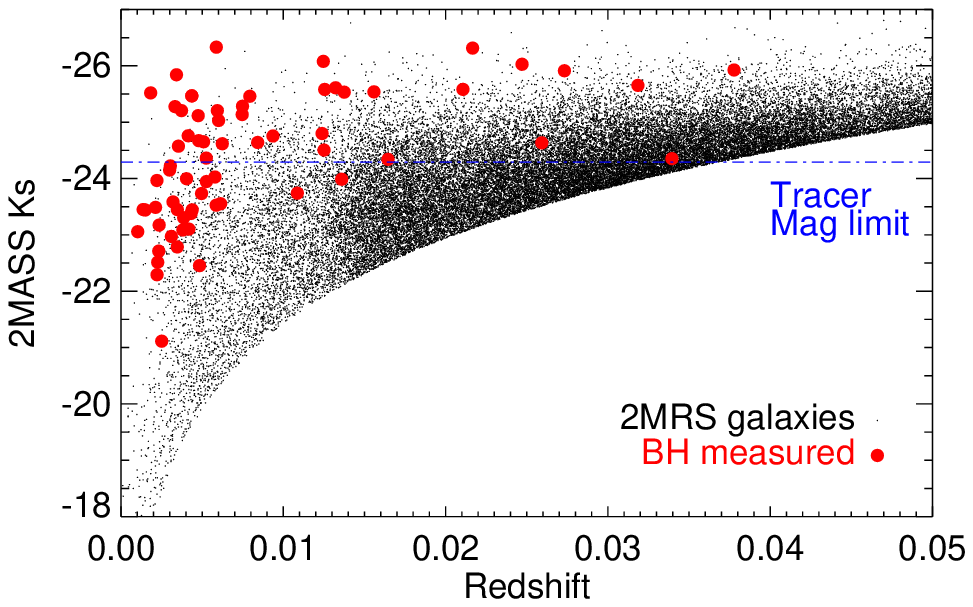}
\caption{Magnitude vs redshift of the 2MASS Redshift Survey (2MRS) and the galaxies with dyanamical black hole measurements. The blue line shows the magnitude limit used for the tracer galaxies in the calculation of the background environmental measures (4th nearest neighbor distance and number density).}
\label{fig_maglimit}
\end{figure}

In an attempt to probe each of the different environments discussed earlier, we will use 4 environmental indicators. Each of these indicators is either calculated from the 2MASS Redshift Survey (2MRS) \citep{Huchra2012} or drawn from previously published analysis by the 2MRS team. The 2MRS was designed to obtain a redshift for all galaxies with Two Micron All Sky Survey \citep[2MASS]{2mass} \Ks\ magnitudes brighter than 11.75 and greater than 5\degree\ from the galactic plane. As mentioned, two galaxies from the original sample of 72 galaxies with good velocity dispersion (\veldisp) and black hole mass (\Mbh) measurements are not in the 2MRS: the Milky Way, for an obvious reason, and Circinus, which is less than 5\degree\ from the galactic plane. Otherwise, 2MRS provides a nearly uniform sampling of the galaxies in the local universe, and thus is ideal for our purposes. While not a strict requirement for our purposes, the \Ks\ selection provides for a nearly stellar mass selected sample. This is particularly useful when measuring environments, where the star formation histories and thus the optical luminosities systematically vary as a function of environment. The distribution of redshift and \Ks\ magnitude of the full 2MRS and of galaxies with good \Mbh\ measurements is shown in Figure \ref{fig_maglimit}. 
 
When calculating any environmental measure from a flux limited survey, care has to be taken to adjust for the varying intrinsic luminosity limit as a function of redshift. One common method is to make a volume limited sample by restricting the tracer galaxies to have absolute magnitude brighter than the flux limit at the maximum redshift of the sample. The most distant galaxy in our sample is the Brightest Cluster Galaxy (BCG) of Abell 1836 at a distance 157.5 \Mpc, corresponding to a absolute magnitude \Ks\ limit of -24.3. This limit is shown as a red line in Figure \ref{fig_maglimit}. This method is used for the first two environmental measures. Alternatively, the magnitude limit can be directly modeled by using a sliding estimate of the typical inter-galaxy distance. This second method was implemented during the creation of the group catalog that we use.

\subsubsection{$N$th nearest neighbor distance}\label{sec-NN}

The projected distance to the nearest Nth galaxy neighbor has been used extensively as an environmental indicator \citep[eg.,][]{dressler80, baldry06, Peng10} and is not strongly dependent on the choice of parameters. To calculate this quantity, we define a subsample of tracer galaxies from the 2MRS which have a magnitude brighter than the \Ks\ limit of -24.3 and a line of sight velocity of $\pm$ $v$ from the target galaxy. The projected distance to the $N$th nearest galaxy in this tracer sample is the quantity of interest. In this paper, we set $v$ = 500 \kms\ and take $N$ = 4. The distribution of the resulting distances is shown in the top panel of Figure \ref{fig_bhenvir}. This is a direct probe of the galaxy density, but is difficult to physically interpret as the effective scale of the measurement varies.

\subsubsection{Fixed aperture number density}\label{sec-ndens}
A complement to the $N$th nearest neighbor distance is a measure of the number of galaxies within a fixed aperture cylindrical volume. This quantity has also been used in many studies \citep[eg.,][]{croton2df, blanton_berlind_07, wilman_multi} and for our purposes is ideal to measure uniformly the very large scale environment of a galaxy. While a typical galaxy cluster may have a virial radius of $\sim$ 2 \Mpc, there are overdensities on much larger scales ($\sim$ 10 \Mpc) which our other indicators do not effectively probe. The calculation of fixed aperture number density requires the definition of a radius, $R$, and line of sight velocity, $\pm$ $v$, to create a cylinder centered on each galaxy. With these definitions, the number of tracer galaxies in each cylinder is counted. In an effort to probe the large scales, we define $R$ as 5 \Mpc\ and $v$ as 1000 km/s. The resulting distribution of number densities is shown in the bottom panel of Figure \ref{fig_bhenvir}.

\subsubsection{Group membership}

The 2MRS team has extensively studied the large scale structure in their survey, presenting both group catalogs \citep{crook_groups} and a reconstructed density field \citep{erdogdu}. We make use of the high density contrast (HDC) group catalog of \citet{crook_groups}, subsequently updated in \citet{crook_groups_update}. This catalog uses the well known friends-of-friends method \citep{huchra} to link galaxies into groups and clusters. In Crook et al., the projected separation linking length, $D$, is scaled with redshift to account for the flux limited sample and thus the varying inter-galaxy distance. In the middle of the redshift range, cz $\sim$ 5000 km/s, the linking length is $\sim$ 1 \Mpc. The line of sight velocity difference is not scaled with redshift and is set to be equal to 350 km/s. \citet{crook_groups} show that this choice of linking parameters corresponds to a density contrast of $\delta \rho/\rho$ = 80. They also present a lower density contrast group catalog (LDC) with larger linking parameters. However, this more generous catalog tends to combine `known' galaxy clusters, such as Abell 426 and Abell 347, into single objects. In this paper we will use the HDC catalog, but we have also done the full analysis using the LDC catalog with no significant change in the results. The \citet{crook_groups} catalog also presents measures of the virial mass of each group. This viral mass, M$_V$, which we also refer to as the halo mass, is a defined as M$_V$ = $\frac{3\pi}{2}$ $\frac{\sigma_p^2 R_p}{G}$, where $\sigma_p^2$ is the projected velocity dispersion of the group,  $R_P$ is the projected virial radius and $G$ is Newton's constant.

There are two possible problems with using the Crook et al. halo masses to compare groups across the redshift range of the 2MRS. Firstly, as the 2MRS survey is flux limited, galaxy groups with the same intrinsic luminiosity distribution will have fewer galaxies within the survey limits at higher redshift. To a first approximation, the Crook et al. algorithim deals with this issue by scaling the linking length with redshift. Nonetheless, galaxy groups which do not have more than three member galaxies above the survey luminosity limit at a given redshift will be missed by the algorithim. In particular, fossil groups or pairs will be missed even if they have the same halo mass as a trio of related galaxies.

Secondly, we must ensure that the halo masses are measured in similar ways regardless of the redshift of the group. As discussed, the halo mass is a result of the velocity dispersion and projected radius, and it is then expected that the galaxies are simply tracers of the underlying mass distribution. However, if galaxy groups experience mass segregation in which higher stellar mass galaxies tracer different regions of the mass distribution than lower stellar mass galaxies, then higher redshift groups could have systematically different halo mass measurements. Luckily, in a large sample of low redshift galaxy clusters \citet{vonderlinden} have found no evidence for mass segregation once the central galaxy was removed. In summary, while the halo mass measurement presented here is inherently noisy, there is no reason to believe it can not be fairly compared across the redshift range. However, we caution that the legitemate galaxy groups may be missed, and thus the most robust comparison of samples split by halo mass would compare 'massive' halos vs the full sample.

\begin{figure}
\leavevmode \epsfysize=8.6cm \epsfbox{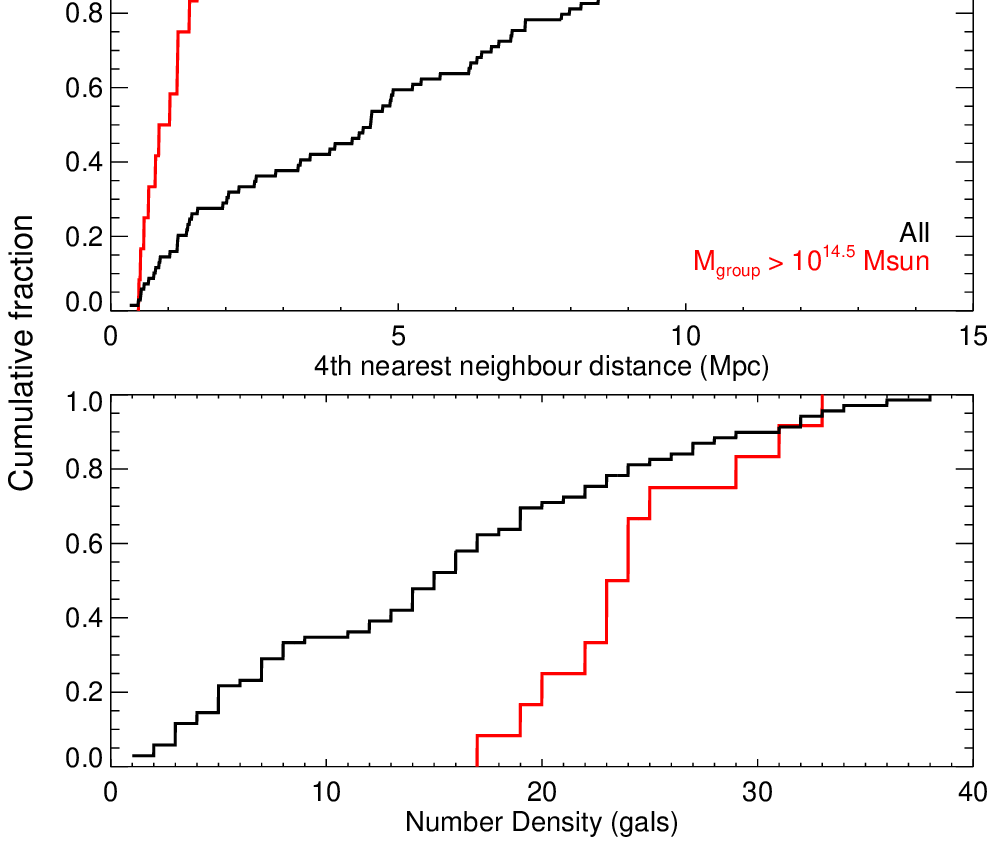}
\caption{The cumulative distributions of environmental indicators. ({\it top}) Shown for the 4th nearest neighbor distance as defined in \textsection \ref{sec-NN}. Dense environments have low nearest neighbor distance.  ({\it bottom}) Shown for the 10 Mpc number density as defined in \textsection \ref{sec-ndens}. In this case, dense environments are characterized by high number density. In both cases, the black lines are all 69 galaxies in our BH sample, while the red lines show only the distribution of the galaxies in galaxy groups with 10$^{14.5}$\Mdot and greater.}
\label{fig_bhenvir}
\end{figure}

 As an example of how the group catalog correlates with our previous measurements, the red line in Figure \ref{fig_bhenvir} represents the distribution of environmental measures for galaxies in halos with total virial masses greater than 10$^{14.5}$\Mdot. Clearly, galaxies in these dense environments are equally well described by any of our measures so far. The 2MRS group catalog is complete only to cz $<$ 10000 \kms, which leaves three galaxies beyond the limits of the catalog (NGC6086, NGC6264 and the BCG of Abell 1836). For the sake of consistency, we remove these galaxies from consideration in this section only. 20 galaxies of the remaining sample of 66 galaxies are not linked to groups in the \citet{crook_groups} catalogue. For these galaxies we will assume that their halo masses are below the limits of detection, and are thus in low mass halos. These galaxies will all be in the lower 50\% samples of group mass. 

\subsubsection{Central or satellite distinction}

As already stated, the location of a galaxy within a dark matter halo plays a key role in its evolution. As a complement to our previous measures, which did not take this into account, we will also calculate whether a galaxy is a central or satellite. For galaxies within \citet{crook_groups}, we will assume that the most luminous galaxy (\Ks\ band) in each group is the central galaxy. While this definition makes no link to the location of a galaxy, it is a common way to divide between centrals and satellites. A key reason for this is the difficulty of measuring the true center of the group of galaxies from their member positions alone. While this may be trivial for a galaxy cluster with a thousand members, it is essentially meaningless for a group of 3 members. Nonetheless, this definition is reasonable, because it picks out the known BCGs of Coma, Abell 3565, Abell 1367, Abell 2162 and Abell 2666. However, NGC1399, which is commonly thought to be the central galaxy in the Fornax cluster, is actually fainter (\Ks\ = 6.4) than a galaxy near the edge of the cluster (NGC1316; \Ks\ = 5.9). Similarly, M87 (\Ks\ = 5.9) is dynamically at the center of the Virgo cluster, but M49 (\Ks\ = 5.5), a galaxy at the center of an infalling subgroup, is actually the brighter galaxy. For consistency with groups which do not have well measured centers, we do not change these. Thus, it is important to remember that our definition of a central is based on luminosity alone. 

For galaxies which are not members of groups, the determination of their central/satellite status is more difficult. While the temptation is to claim that all non-grouped galaxies are centrals, this is clearly not the case. We will assume that a galaxy is a central if it is the brightest \Ks\ galaxy within a given radius and velocity range. One option for the radius to use would be to scale by the virial radius of a typical halo hosting a galaxy of the given magnitude. However, while the virial radius of a typical low mass galaxy is on the order of a few hundred kpcs, a galaxy cluster may be 2 \Mpc. So, a galaxy may appear to be a central even while being embedded in a massive cluster with a much more luminous BCG. 

\citet{geha_stellar} have examined the environments of low mass galaxies (10$^7$-10$^9$\Mdot) and found that the fraction of quenched galaxies, defined as having no apparent H$\alpha$ emission and strong 4000$\AA$ breaks, is extremely small ($<$ 0.06$\%$) when the distance to the nearest massive galaxy is more than 1.5 \Mpc. In contrast, the fraction is $\sim$ 1-2 $\%$ at a distance of 1 \Mpc\ and $\sim$ 10-15 $\%$ at 0.5 \Mpc. This suggests that a reasonable value for the radius would be 1.5 \Mpc. For the velocity difference we are motivated by the 350 \kms\ velocity difference used in the creation of the group catalog. Therefore, we call all non-grouped galaxies which have a more luminous \Ks\ band galaxy less than 1.5 Mpc and $\pm$ 350 km/s away a satellite galaxy. This adds an additional 6 satellites from the population of 23 non-grouped galaxies. 

 The median 4th nearest neighbour distance of galaxies in the 2MRS is 4.83 Mpc, whereas the sample of galaxies with BH measurements have a median 4th nearest neighbour distance indicative of slightly denser environments (4.53 Mpc). Similarly, the satellite fraction of 42$\%$ (29/69), is higher than the $\sim$ 30 $\%$ found in large scale surveys \citep{zehavi2005occupation,zheng2007occupation}. Thus, the average environment of galaxies with black hole measurements may be slightly denser than the typical galaxy. However, the satellite fraction is expected to be a strong function of stellar mass and colour, where low mass ($<$ 10$^9$ \Msun) red galaxies have satellite fractions of $\sim$ 60 $\%$ \citep{yang2008occupation}.

\subsection{Line fitting}

We parameterizing the power law scaling relations between black hole mass and other parameters, following \citet{McConnellMa2013}, as:

\begin{equation}\label{equ-scaling}
\mathrm{log}_{10} M_{\bullet} = \alpha + \beta \, \mathrm{log}_{10} X \, , 
\end{equation} where $X$ is either $\sigma$/200\kms, $L_V$/10$^{11}$\Lsun\ or $M_{bulge}$/10$^{11}$\Msun\ and \Mbh\ is measured in units of \Msun. Thus, the slope and intercept of each scaling relation is given by $\beta$ and $\alpha$ respectively.

We fit this equation using the linear regression Bayesian estimator of \citet{Kelly_linmixerr}. This algorithm accounts for measurement error on both variables and also produces an estimate of the intrinsic scatter, $\epsilon$. \citet{McConnellMa2013} found this method gave consistent results with the least squares estimator ${\tt MPFITEXY}$ \citep{williams_mpfit} which builds on ${\tt MPFIT}$ \citep{mpfit}. There has been much discussion in the literature about how to fit the power law scaling relations. While this is critical for examining the absolute values of the scaling relations, our goal is to examine the differential evolution of subdivided samples and is thus slightly less dependent on the details of the regression scheme. We perform the linear regressions in the forward direction, but we note that the qualtitive trends we see are similar if the linear regression is done in the inverse direction. 

\section{Results} \label{sec-results}

We will now examine the results of performing linear regressions on each of the three primary black hole scaling relations while dividing into several subsamples based on the calculated environmental indicators. 

\subsection{\Mbh -$\sigma$ relation}

First, we look at the full sample of galaxies with which we were able to make good environmental measurements, a total sample of 69 galaxies. We find $\alpha$ = 8.33 $\pm$ 0.05 and $\beta$ = 5.72 $\pm$ 0.33 with a scatter, $\epsilon$, of 0.37 $\pm$ 0.04. As a comparison, \citet{McConnellMa2013} find very similar results of ($\alpha$, $\beta$, $\epsilon$) = (8.32 $\pm$ 0.06, 5.58 $\pm$ 0.34, 0.42 $\pm$ 0.04) using the same fitting method, but with the full sample of 72 galaxies. Our removal of 3 galaxies has not affected the bulk properties of the scaling relation, and we confirm that we obtain exactly the same results as \citet{McConnellMa2013} for the full sample. 

\begin{figure}
\leavevmode \epsfysize=20.6cm \epsfbox{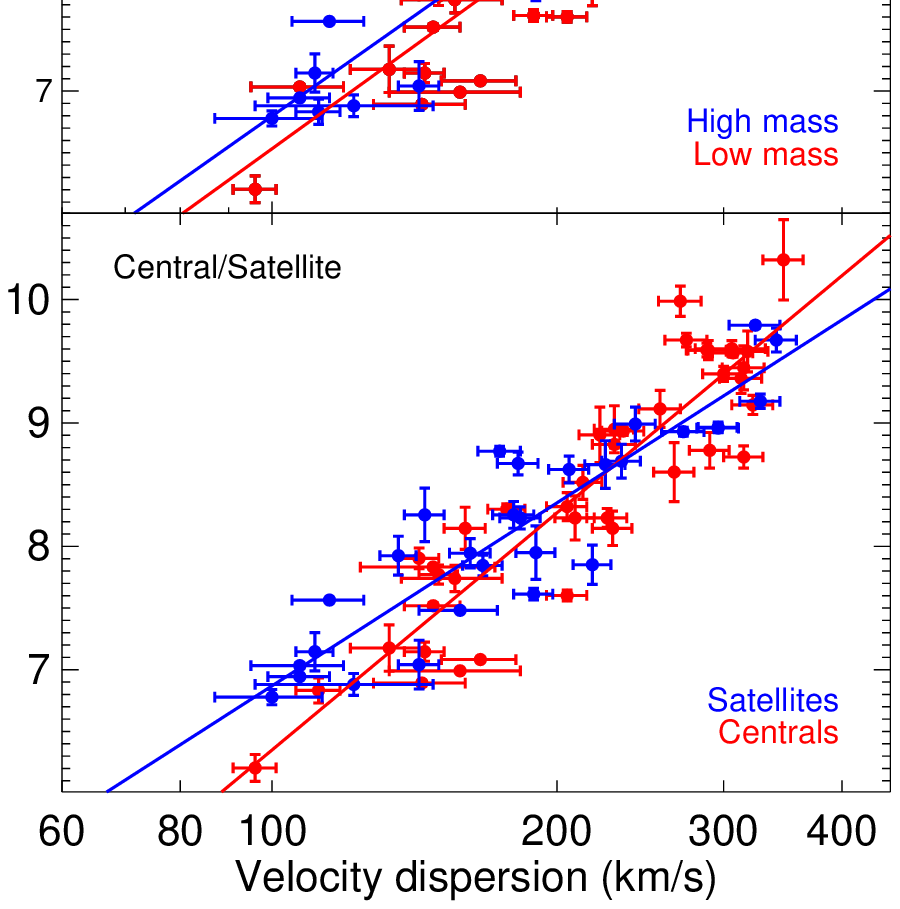}
\caption{\Mbh\ - $\sigma$ relation of environmentally defined samples of galaxies. The sample is divided in two, based on the 4th nearest neighbour distance ({\it top}), number of galaxies within 5 \Mpc\ and $\pm$ 500 \kms\ ({\it second from top}),  the group virial mass ({\it third from top}) and whether the galaxy is a central or satellite ({\it bottom}). The first three indicators are divided by the midpoint in the distribution. In these cases, galaxies in dense environments are blue and those in isolated environments are red. The lines show the linear regression fits of each subsample.
}
\label{fig_bhmsigma}
\end{figure}

We sort the 69 galaxies by each of 4th nearest neighbor, 5 \Mpc\ number density and total halo mass, and perform a linear regression using the lowest 25$\%$, the lowest 50$\%$, the highest 50$\%$ and the highest 25$\%$ of each indicator. While we call the percentiles by these names, in detail, depending on the number of galaxies, the exact percentiles vary slightly throughout the sample. While breaking up the sample into these percentiles is ad hoc, it is based on the studies that find roughly half of typical galaxies are located within galaxy groups or pairs \citep{2pigg, berlind}. As such, we focus on analysising the lowest 50$\%$ and highest 50$\%$ categories, but we emphasis that none of the conclusions would change if instead we focussed on only the upper 25$\%$ and lower 25$\%$ categories. The full results, including the intercept, slope and scatter, as well as the number of galaxies in each subsample and the exact environmental divisions, are presented in Table \ref{tab_main}.

The top three panels of Figure \ref{fig_bhmsigma} show the results of fitting the scaling relations to the highest 50$\%$ and lowest 50$\%$ of each of 4th nearest neighbor, 5 \Mpc\ number density and group mass. In each case, the intercept, $\alpha$, of the relation in the dense environment (ie, shortest 4th nearest neighbour distance, highest number density, and largest halo mass) is systematically higher than in the more isolated environments. The largest difference occurs in the halo mass division, where $\alpha$ is 8.20 $\pm$ 0.08 in low mass halos and 8.43 $\pm$ 0.07 in high mass halos. The slopes of the relations are essentially indistinguishable in either environment. For instance, in more massive halos, the slope is $\beta$ = 5.38 $\pm$ 0.62, and in smaller halos it is 5.57 $\pm$ 0.62. The intrinsic scatter in each subsample is in the range $\epsilon$ = 0.35 - 0.48 without a significant systematic trend with environment. 

In the bottom panel of Figure \ref{fig_bhmsigma}, we show the results of separately fitting the satellite galaxies and the central galaxies. Interestingly, here the results are different from the other environmental indicators. The central and satellite galaxies have similar intercepts $\alpha$ (8.28 $\pm$ 0.07 vs 8.35 $\pm$ 0.08) but different slopes. In the sample of central galaxies $\beta$ = 6.38 $\pm$ 0.49 while in satellites $\beta$ = 4.91 $\pm$ 0.49. The departure from the trends of the other indicators makes sense. At high velocity dispersion,  central galaxies will be in massive halos, while at low velocity dispersion, satellites will be in more massive halos. The steep slope of the central galaxy relation leads to a vertically offset relation when dividing only be group mass.

\begin{table*}
\begin{tabular}{lcccc}
\hline 
Sample & N$_{gal}$ & $\alpha$ & $\beta$ & $\epsilon$ \\ 
\hline
\multicolumn{5}{| c |}{M$_\bullet$ - $\sigma$ relation - Full Sample}\\
\hline
All Galaxies  & 69 &      8.33 $\pm$      0.05 &      5.72 $\pm$      0.33 &      0.37 $\pm$      0.04 \\ 
4th Nearest - lowest 25 \% (N$_4$ $\leq$ 1.38 Mpc)  & 17 &      8.40 $\pm$      0.12 &      5.44 $\pm$      0.77 &      0.48 $\pm$      0.13 \\ 
4th Nearest - lowest 50 \% (N$_4$ $<$ 4.53 Mpc)  & 34 &      8.39 $\pm$      0.08 &      5.68 $\pm$      0.45 &      0.40 $\pm$      0.06 \\ 
4th Nearest - highest 50 \% (N$_4$ $\geq$ 4.53 Mpc)  & 35 &      8.26 $\pm$      0.07 &      5.76 $\pm$      0.56 &      0.37 $\pm$      0.06 \\ 
4th Nearest - highest 25 \% (N$_4$ $\geq$ 6.99 Mpc)  & 18 &      8.10 $\pm$      0.13 &      4.75 $\pm$      1.14 &      0.46 $\pm$      0.11 \\ 
Number Density - lowest 25 \% (N$_N$ $\leq$ 7)  & 20 &      8.23 $\pm$      0.10 &      5.94 $\pm$      0.85 &      0.40 $\pm$      0.10 \\ 
Number Density - lowest 50 \% (N$_N$ $<$ 15)  & 33 &      8.23 $\pm$      0.07 &      6.00 $\pm$      0.57 &      0.35 $\pm$      0.06 \\ 
Number Density - highest 50 \% (N$_N$ $\geq$ 15)  & 36 &      8.41 $\pm$      0.08 &      5.74 $\pm$      0.46 &      0.41 $\pm$      0.07 \\ 
Number Density - highest 25 \% (N$_N$ $\geq$ 22)  & 19 &      8.45 $\pm$      0.10 &      6.12 $\pm$      0.64 &      0.36 $\pm$      0.11 \\ 
Halo Mass - lowest 50 \% (M$_{halo}$ $<$ 13.29 \Msun)  & 31 &      8.20 $\pm$      0.08 &      5.57 $\pm$      0.62 &      0.40 $\pm$      0.07 \\ 
Halo Mass - highest 50 \% (M$_{halo}$ $\geq$13.29 \Msun)  & 35 &      8.43 $\pm$      0.07 &      5.38 $\pm$      0.44 &      0.36 $\pm$      0.07 \\ 
Halo Mass - highest 25 \% (M$_{halo}$ $\geq$14.02 \Msun)  & 17 &      8.48 $\pm$      0.09 &      5.74 $\pm$      0.49 &      0.33 $\pm$      0.10 \\ 
Satellites & 29 &      8.35 $\pm$      0.08 &      4.91 $\pm$      0.49 &      0.35 $\pm$      0.06 \\ 
Centrals  & 40 &      8.28 $\pm$      0.07 &      6.38 $\pm$      0.49 &      0.38 $\pm$      0.06 \\ 
\hline
\multicolumn{5}{| c |}{M$_\bullet$ - $\sigma$ relation - Early Type galaxies}\\
\hline
All Galaxies  & 52 &      8.37 $\pm$      0.06 &      5.47 $\pm$      0.45 &      0.37 $\pm$      0.05 \\ 
4th Nearest - lowest 25 \% (N$_4$ $\leq$ 1.35 Mpc)  & 13 &      8.51 $\pm$      0.14 &      5.23 $\pm$      0.95 &      0.49 $\pm$      0.18 \\ 
4th Nearest - lowest 50 \% (N$_4$ $<$4.54 Mpc)  & 25 &      8.50 $\pm$      0.10 &      5.07 $\pm$      0.63 &      0.40 $\pm$      0.08 \\ 
4th Nearest - highest 50 \% (N$_4$ $\geq$4.54 Mpc)  & 27 &      8.28 $\pm$      0.09 &      5.66 $\pm$      0.70 &      0.37 $\pm$      0.07 \\ 
4th Nearest - highest 25 \% (N$_4$ $\geq$6.76 Mpc)  & 14 &      8.17 $\pm$      0.15 &      5.90 $\pm$      1.61 &      0.48 $\pm$      0.16 \\ 
Number Density - lowest 25 \% (N$_N$ $\leq$ 7)  & 14 &      8.16 $\pm$      0.13 &      6.72 $\pm$      1.23 &      0.36 $\pm$      0.13 \\ 
Number Density - lowest 50 \% (N$_N$ $<$15)  & 25 &      8.16 $\pm$      0.10 &      6.55 $\pm$      0.83 &      0.34 $\pm$      0.07 \\ 
Number Density - highest 50 \% (N$_N$ $\geq$15)  & 27 &      8.49 $\pm$      0.09 &      5.30 $\pm$      0.60 &      0.41 $\pm$      0.08 \\ 
Number Density - highest 25 \% (N$_N$ $\geq$22)  & 15 &      8.53 $\pm$      0.10 &      5.63 $\pm$      0.67 &      0.34 $\pm$      0.11 \\ 
Halo Mass - lowest 50 \% (M$_{halo}$ $<$13.40 \Msun)  & 23 &      8.28 $\pm$      0.10 &      5.02 $\pm$      0.78 &      0.42 $\pm$      0.08 \\ 
Halo Mass - highest 50 \% (M$_{halo}$ $\geq$13.40 \Msun)  & 27 &      8.46 $\pm$      0.09 &      5.20 $\pm$      0.65 &      0.37 $\pm$      0.07 \\ 
Halo Mass - highest 25 \% (M$_{halo}$ $\geq$14.16 \Msun)  & 13 &      8.43 $\pm$      0.14 &      5.95 $\pm$      0.85 &      0.40 $\pm$      0.16 \\ 
Satellites & 23 &      8.40 $\pm$      0.09 &      4.85 $\pm$      0.60 &      0.37 $\pm$      0.08 \\ 
Centrals  & 29 &      8.26 $\pm$      0.12 &      6.57 $\pm$      0.83 &      0.40 $\pm$      0.07 \\ 
\hline
\multicolumn{5}{| c |}{M$_\bullet$ - M$_{\mathrm{bulge}}$ relation} \\
\hline
All Galaxies  & 34 &      8.45 $\pm$      0.09 &      1.09 $\pm$      0.17 &      0.39 $\pm$      0.08 \\ 
4th Nearest - lowest 50 \% (N$_4$ $<$ 4.54 Mpc)  & 16 &      8.49 $\pm$      0.21 &      1.01 $\pm$      0.28 &      0.45 $\pm$      0.16 \\ 
4th Nearest - highest 50 \% (N$_4$ $\geq$4.54 Mpc)  & 18 &      8.44 $\pm$      0.13 &      1.19 $\pm$      0.36 &      0.45 $\pm$      0.15 \\ 
Number Density - lowest 50 \% (N$_N$ $<$15)  & 16 &      8.38 $\pm$      0.15 &      1.11 $\pm$      0.28 &      0.44 $\pm$      0.16 \\ 
Number Density - highest 50 \% (N$_N$ $\geq$15)  & 18 &      8.52 $\pm$      0.14 &      1.04 $\pm$      0.21 &      0.46 $\pm$      0.14 \\ 
Halo Mass - lowest 50 \% (M$_{halo}$ $<$ 13.40 \Msun)  & 16 &      8.23 $\pm$      0.13 &      1.07 $\pm$      0.30 &      0.42 $\pm$      0.16 \\ 
Halo Mass - highest 50 \% (M$_{halo}$ $\geq$ 13.40 \Msun)  & 17 &      8.68 $\pm$      0.12 &      1.01 $\pm$      0.21 &      0.36 $\pm$      0.13 \\ 
Satellites & 16 &      8.64 $\pm$      0.13 &      1.01 $\pm$      0.29 &      0.45 $\pm$      0.16 \\ 
Centrals  & 18 &      8.10 $\pm$      0.15 &      1.45 $\pm$      0.23 &      0.29 $\pm$      0.14 \\ 
\hline
\multicolumn{5}{| c |}{M$_\bullet$ - L$_{\mathrm{bulge,V}}$ relation} \\
\hline
All Galaxies & 43 &      9.25 $\pm$      0.11 &      1.13 $\pm$      0.17 &      0.53 $\pm$      0.07 \\ 
4th Nearest - lowest 50 \% (N$_4$ $<$ 4.39 Mpc)  & 21 &      9.17 $\pm$      0.14 &      1.38 $\pm$      0.29 &      0.57 $\pm$      0.11 \\ 
4th Nearest - highest 50 \% (N$_4$ $\geq$4.39 Mpc)  & 22 &      9.35 $\pm$      0.19 &      1.10 $\pm$      0.25 &      0.55 $\pm$      0.10 \\ 
Number Density - lowest 50 \% (N$_N$ $<$15)  & 20 &      9.29 $\pm$      0.16 &      0.96 $\pm$      0.24 &      0.52 $\pm$      0.12 \\ 
Number Density - highest 50 \% (N$_N$ $\geq$15)  & 23 &      9.19 $\pm$      0.15 &      1.29 $\pm$      0.25 &      0.59 $\pm$      0.11 \\ 
Halo Mass - lowest 50 \% (M$_{halo}$ $<$ 13.46 \Msun)  & 20 &      9.17 $\pm$      0.18 &      1.29 $\pm$      0.27 &      0.51 $\pm$      0.12 \\ 
Halo Mass - highest 50 \% (M$_{halo}$ $\geq$ 13.46 \Msun)  & 21 &      9.31 $\pm$      0.17 &      0.94 $\pm$      0.24 &      0.62 $\pm$      0.12 \\ 
Satellites & 18 &      9.38 $\pm$      0.29 &      1.09 $\pm$      0.33 &      0.63 $\pm$      0.15 \\ 
Centrals  & 25 &      9.18 $\pm$      0.11 &      1.46 $\pm$      0.24 &      0.49 $\pm$      0.10 \\ 
\hline
\end{tabular}
\label{tab_main}

\caption{The results of linear regression of the black hole scaling relations for environmentally defined subsamples of galaxies. The scaling relations are given by Equation \ref{equ-scaling}, and are presented for \Mbh-$\sigma$ (the full sample and after restricting to only early type galaxies), \Mbh-M$_{V}$ and \Mbh-M$_{\mathrm{bulge}}$. $\epsilon$ is an estimate of the intrinsic scatter in the relation, while $\alpha$ and $\beta$ are the intercept and slope, respectively. The regressions are performed with the ${\tt linmix \_ err}$ method of \citet{Kelly_linmixerr}.}
\label{tab_main}
\end{table*}

\subsubsection{Early types only} 

\begin{figure}
\leavevmode \epsfysize=20.6cm \epsfbox{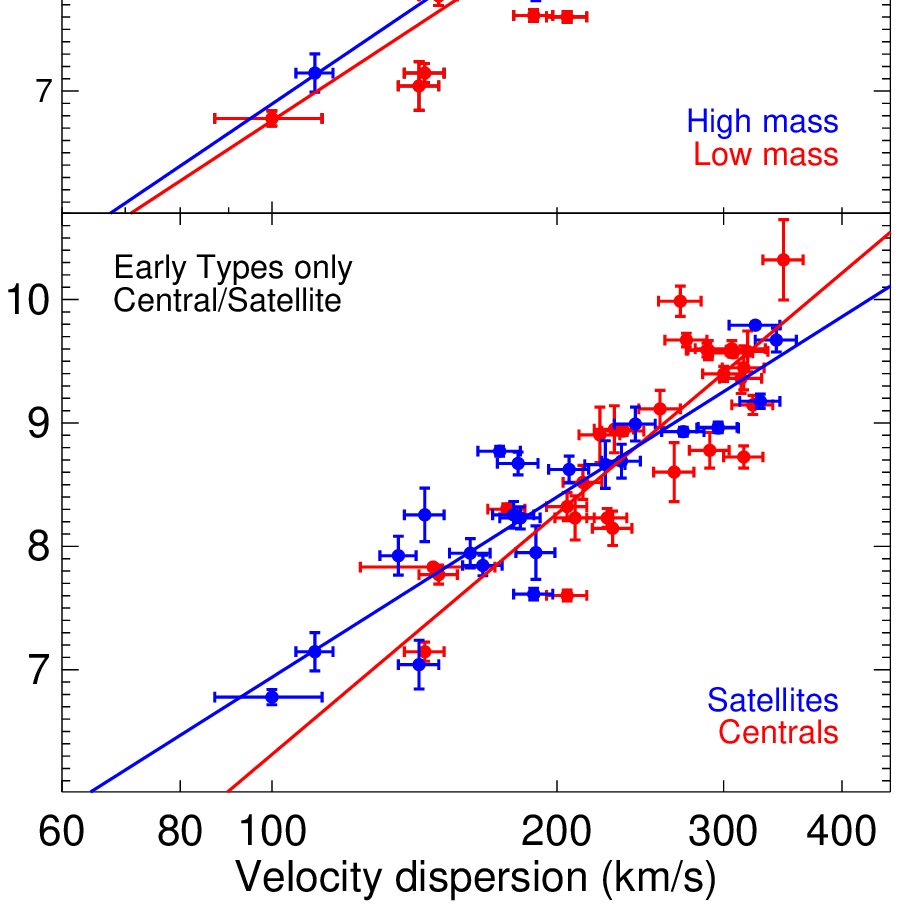}
\caption{\Mbh\ - $\sigma$ relation for environmentally defined samples of early type galaxies. The sample is divided in two, based on the 4th nearest neighbor distance ({\it top}), number of galaxies within 5 \Mpc\ and $\pm$ 500 \kms\ ({\it second from top}),  the group virial mass ({\it third from top}) and whether the galaxy is a central or satellite ({\it bottom}). The first three indicators are divided by the midpoint in the distribution. In these cases, galaxies in dense environments are blue and those in isolated environments are red. The lines show the linear regression fits of each subsample. }
\label{fig_bhmsigma_early}
\end{figure}

Because of the known morphology-density relation, it is possible the observed environmental dependence of the scaling relations is created by the underlying morphology differences and the change in relative fractions between environments. Consequently, we restrict only to those galaxies classified as early types (either E or S0) by \citet{McConnellMa2013}. The intention of this selection is to remove spiral galaxies from consideration, as the fraction of these galaxies strongly varies with environment \citep{dressler80, postman, mcgee, Bamford_zoo}. This reduces the sample to 52 total galaxies. We redo the linear regressions after redividing the sample by the distributions of their environmental indicators. The results of these regressions are shown in the second section of Table \ref{tab_main} and are plotted in Figure \ref{fig_bhmsigma_early}. 

Most interestingly, the relations for 4th nearest neighbor, group mass and central/satellite give no qualitative change. Indeed, both the intercept, $\alpha$, and the slope, $\beta$, of the relations for the central and satellite subsamples are nearly identical whether all galaxies are considered or only the early types. Clearly, the environmental dependence of the \Mbh\ - $\sigma$ relation does not appear to be coming simply from the changing mixture of early type and late type with environment. However, this environmental dependence may still reflect finer subsamples of early and late type galaxies. In particular, \citet{Graham2008a, Graham2008b} have shown that barred galaxies have considerable offset from the \Mbhrel\ relation of non-barred galaxies. 

The only significant change between the full sample and the early type sample is the increasing $\beta$ in the case of high number density, accompanied by a reduction in $\beta$ with low number density. Nonetheless, the early type value of each slope is within the uncertainty of the slope determined from the full sample.

\subsubsection{The role of BCGs} 

\begin{figure}
\leavevmode \epsfysize=7.6cm \epsfbox{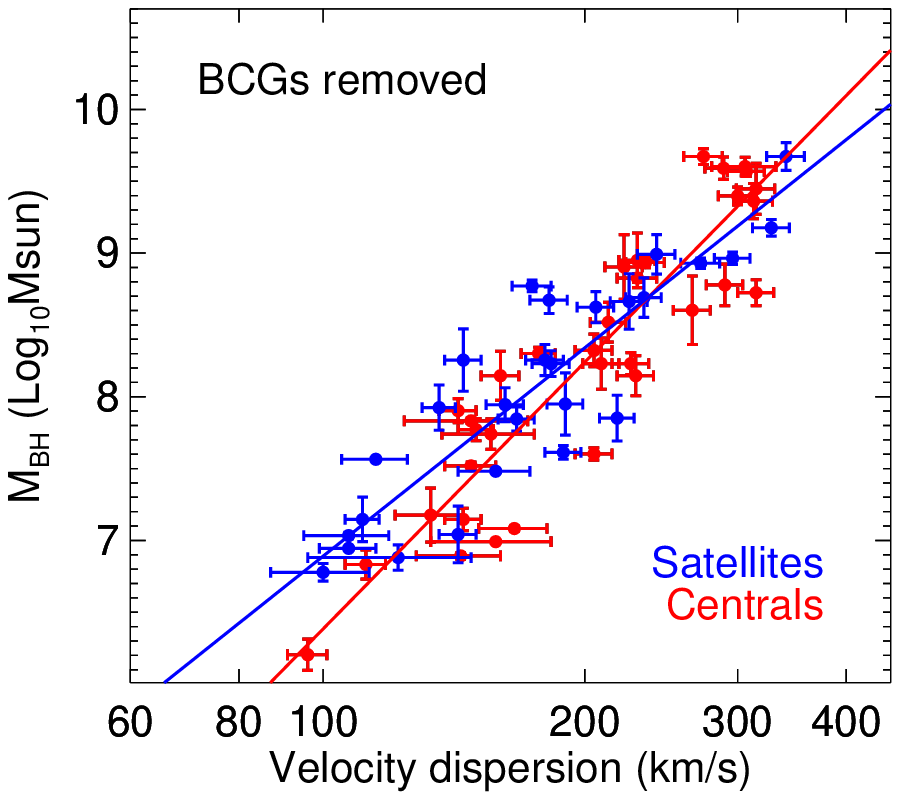}
\caption{\Mbh\ - $\sigma$ relation for central and satellite galaxies after the removal of the 8 known BCGs. The blue and red lines show the results of a linear regression on the centrals and satellites respectively. }
\label{fig_bhmsigma_nobcg}
\end{figure}

Until now we have treated all central galaxies the same way. However, there is some reason to expect that central galaxies of massive clusters have fundamentally different formation mechanisms from normal, isolated central galaxies. To test to what extant these BCGs are creating the different scaling relations, we will remove them from the sample. The BCGs, as defined by \citet{McConnellMa2013}, are the BCGs of Abell 1836 and 3565 as well as NGC1399, NGC3842, NGC4486 (M87), NGC4889, NGC6086 and NGC7768. Again, we redivide the sample by the distributions of environmental indicators and perform linear regressions to each scaling relation. The results are shown in Figure \ref{fig_bhmsigma_nobcg} and are tabulated in Table \ref{tab_nobcg}. The intercepts of the regression have no significant change after the removal of the BCGs. In contrast, the slope of the centrals becomes shallower by $\sim$ 0.2 dex after the removal of the BCGs. However, there is still a significant difference in the slope of the relations between satellites and centrals after the removal of the known BCGs.

\subsection{\Mbh - $M_{bulge}$ relation}\label{sec-mbhbulge}

\begin{figure}
\leavevmode \epsfysize=20.6cm \epsfbox{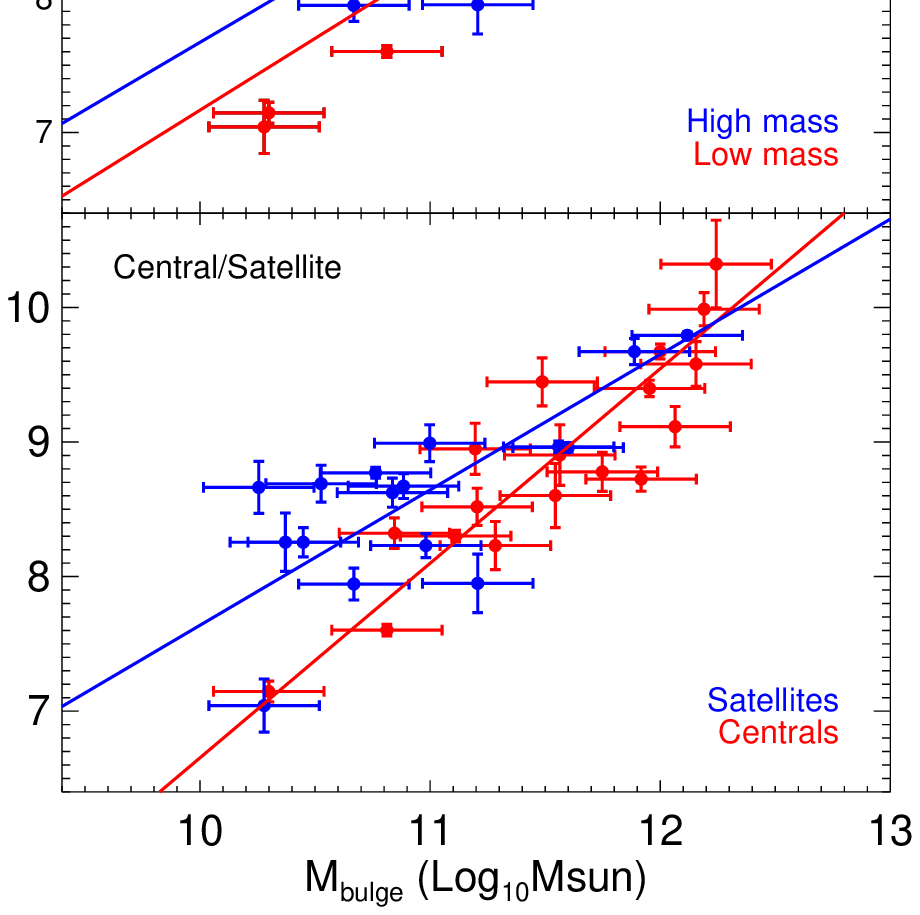}
\caption{Mbh\ - M$_\mathrm{bulge}$ relation of environmentally defined samples of galaxies. The sample is divided in two based on the 4th nearest neighbor distance ({\it top}), number of galaxies within 5 \Mpc\ and $\pm$ 500 \kms\ ({\it second from top}), the group virial mass ({\it third from top}) and whether the galaxy is a central or satellite ({\it bottom}). The first three indicators are divided by the midpoint in the distribution. In these cases, galaxies in dense environments are blue and those in isolated environments are red. The lines show the linear regression fits of each subsample.}
\label{fig_bhmbulge}
\end{figure}

The \Mbhbulge\ relation is potentially a powerful indicator of the formation history of bulges. In this section, we examine this relation for the 34 galaxies in the sample with good dynamical measurements of bulge mass. This sample includes only early type galaxies. Using the full sample of 34 galaxies, we find relations of $\alpha$ = 8.45 $\pm$ 0.09 and $\beta$ = 1.09$\pm$ 0.17, which are similar to the results of \citet{McConnellMa2013}. Performing a linear regression on several environmentally divided samples using Equation \ref{equ-scaling} gives the results tabulated in the third panel of Table \ref{tab_main}. The fewer number of galaxies in this sample prohibits finer bins than dividing in half. We plot the results, including the determined regression lines, in Figure \ref{fig_bhmbulge}. 

Dividing the sample into isolated and dense environments via 4th nearest neighbor or 5 \Mpc\ number density leads to no statistically significant difference in the scaling relation. In contrast, we see that the group mass division leads to an increase in $\alpha$ in dense environments. The most massive halos ($\alpha$ = 8.68 $\pm$ 0.12) tend to host BHs which are higher for a given bulge mass than is the case for lower mass halos ($\alpha$ = 8.23 $\pm$ 0.13). Despite this offset, the slopes of the two relations are indistinguishable. 

The central galaxies ($\beta$ = 1.45 $\pm$ 0.23) have much steeper slope than the satellite galaxies ($\beta$ = 1.01 $\pm$ 0.29) but also have a lower intercept ($\alpha$ = 8.10 $\pm$ 0.15 vs 8.64 $\pm$ 0.13). Unfortunately, in addition to the small sample size, the restriction to a sample of early type galaxies results in the centrals and satellites populating different sections of the \Mbhbulge\ relation. While most of the central galaxies in this plot have bulge masses greater than 10$^{11}$ \Mdot\, there are many satellite galaxies below that mass. This is not unexpected, as it is well known that low mass early type galaxies are preferentially satellite galaxies, but it makes it difficult to determine what is driving the difference in this sample. Restricting to a mass matched sample is not possible with the small numbers. 

\subsection{\Mbh - $L_{\mathrm{bulge,V}}$ relation}

\begin{figure}
\leavevmode \epsfysize=20.6cm \epsfbox{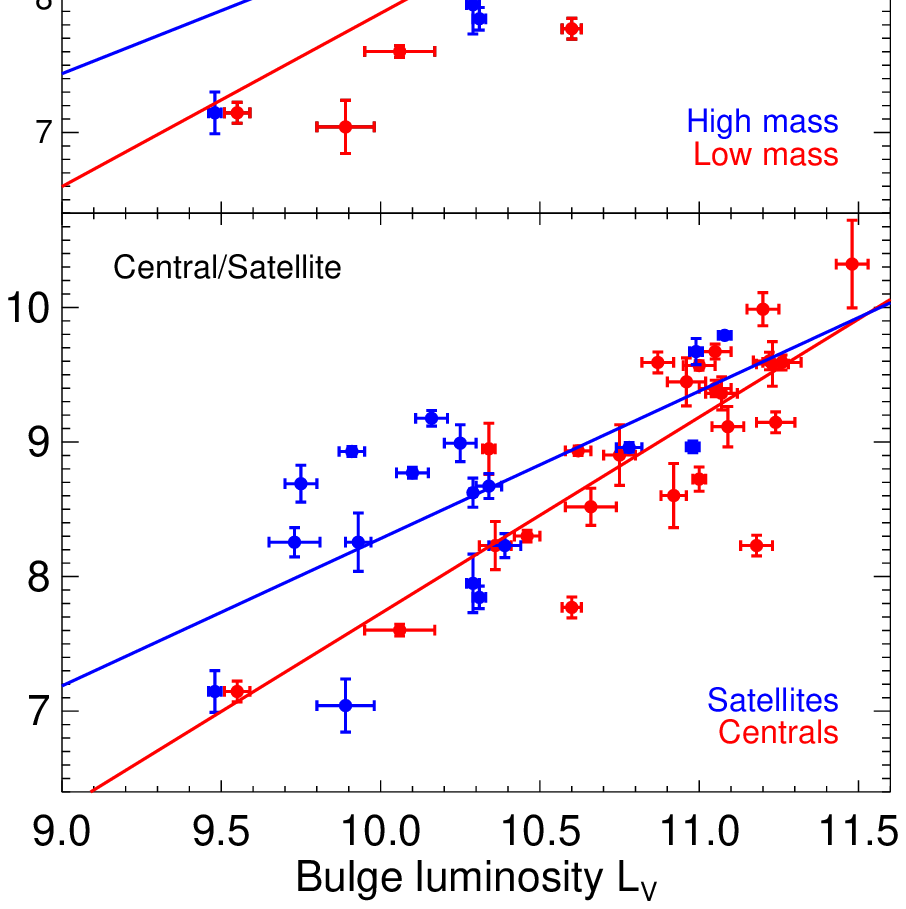}
\caption{Mbh\ - L$_{\mathrm{V,bulge}}$ relation of environmentally defined samples of galaxies. The sample is divided in two based on the 4th nearest neighbor distance ({\it top}), number of galaxies within 5 \Mpc\ and $\pm$ 500 \kms\ ({\it second from top}),  the group virial mass ({\it third from top}) and whether the galaxy is a central or satellite ({\it bottom}). The first three indicators are divided by the midpoint in the distribution. In these cases, galaxies in dense environments are blue and those in isolated environments are red. The lines show the linear regression fits of each subsample.}
\label{fig_bhmluminosity}
\end{figure}

Another often used BH scaling relation is that between the BH mass and the luminosity of the bulge. This is similar to using the bulge mass, except it includes the variation of mass to light ratios (M/L) of the bulges which can encode information about metallicity, star formation history, dust, etc. Again, we use the sample of \citet{McConnellMa2013} for the bulge luminosities of 43 galaxies in the V band. Our regression of the full sample ($\alpha$ = 9.25$\pm$0.11, $\beta$ = 1.13$\pm$0.17) is consistent with the \citet{McConnellMa2013} results. The results for environmentally split samples are shown in Figure \ref{fig_bhmluminosity} and tabulated in the bottom panel of Table \ref{tab_main}. 

The dense environments (as measured by 4th nearest neighbor or number density), the more massive groups, and central galaxies all have steeper slopes and lower intercepts than their corresponding samples. However, before interpreting these results too deeply, we note that the difference between the luminosity distributions of the environmentally split samples is even more extreme than the bulge mass distributions of the central/satellite relation mentioned in \textsection \ref{sec-mbhbulge}. Satellite galaxies are more likely to have truncated star formation, so they are likely to have lower M/L than central galaxies. Thus a central and satellite galaxy may have the same bulge mass, but recent star formation in the central would make it brighter. If we assume that the power law in Equation \ref{equ-scaling} is a perfect description of the data, then we should not be concerned by the different distributions. However, \citet{Graham2012} has shown that the bulge mass and bulge luminosity scaling relations are not well described by single power laws. Indeed, it seems clear that the much better studied velocity dispersion - luminosity relationship is not a single power law, which implies that the \Mbhrel\ and \mbhbullum\ can not both be power laws \citep{Binney82, matkovic05, Graham_review}. It is interesting that we do see a significant difference between each of the environmentally split samples; however, to tell if this relates to the different luminosity distributions and an underlying non-power law relation, or an intrinsic environmental difference, requires a more extended sample. Given that high luminosity satellites are unlikely, the best progress can be made by examining a sample of early type, low luminosity centrals (if they exist). 

\begin{table}
\begin{tabular}{lcccc}
\hline 
Sample & N$_{gal}$ & $\alpha$ & $\beta$ & $\epsilon$ \\ 
\hline
Satellites & 27 &      8.34 $\pm$      0.08 &      4.81 $\pm$      0.54 &      0.37 $\pm$      0.08 \\ 
Centrals  & 34 &      8.24 $\pm$      0.07 &      6.15 $\pm$      0.52 &      0.37 $\pm$      0.07 \\ 
\hline
\end{tabular}
\label{tab_nobcg}

\caption{The results of linear regression of the black hole scaling relations for central and satellite galaxies after removing the known massive BCG galaxies.  The table lists the number of galaxies in each sample as well as the intercept ($\alpha$), slope ($\beta$), and intrinsic scatter ($\epsilon$).}
\label{tab_nobcg}
\end{table}

\section{Discussion}\label{sec-discuss}

In each of the three scaling relations we have examined, central galaxies have shown a steeper slope than satellite galaxies. The satellite galaxies in the \Mbhbulge\ relation also showed a higher intercept than the central galaxies, while the differences in the intercepts of the other two relations were not statistically significant. The difference in the scaling relations between central and satellite galaxies was consistently the most significant of any of the environmental divisions. Further, this difference was maintained after restricting to early type only samples or removing the known BCGs. For these reasons, we will assume that the fundamental difference in environmentally divided scaling relations is driven by the difference between the central and satellite galaxies in the \Mbhrel\ relation. While the discussion below is framed in those terms, the focus on this instead of another relation is largely because velocity dispersion is a direct probe of the potential well and thus the needed escape velocity. We emphasize that the absence of evidence for strong environmental signatures in the scaling relations when divided by nearest neighbour distance, number density or halos mass does not imply there is no such effect. It may be that the environmental divisions used are too coarse, or more data is needed. We focus on the central/satellite division because of the strong effect it shows.

\subsection{Secular growth of black holes}

Many authors have investigated the origins of the \Mbhrel\ relation analytically by assuming that the \Mbh\ mass is due to the secular growth of a galaxy without continued cosmological accretion. \citet{silk_rees} have shown that an AGN driven outflow which conserves its energy will result in an \Mbhrel\ relation of the form \Mbh $\propto$ $\sigma^5$.  \citet{King03} pointed out that because of the short Compton cooling time in the inner regions of the galaxy, AGN wind shocks would quickly cool and the outflow would be `momentum-driven'. He found that the momentum driven wind leads to a relation of the form \Mbh $\propto$ $\sigma^4$. After including the adjustment for sub-Eddington luminosities suggested by \citet{Murray05}, the full relation is

\begin{equation}\label{equ-msigma}
M_\sigma = \frac{f_g\ \kappa}{\pi G^2 \Gamma}\sigma^4
\end{equation}
where $f_g$ is the gas fraction, $\kappa$ is the opacity and $\Gamma$ is the fraction of the Eddington luminosity equal to the AGN luminosity ($\frac{L_{\mathrm{BH}}}{L_{\mathrm{Edd}}}$). To explain the differential slopes of the centrals and satellites requires an additional dependence on $\sigma$ for the centrals but not the satellites. The slope of satellites is roughly consistent with a M$_{BH}$ $\propto$ $\sigma^4$. Thus, the slopes imply that $\frac{f_g\ \kappa}{\Gamma}$ $\propto$ $\sigma^\beta$ where $\beta \sim$ 2 for centrals and $\sim$ 0 for satellites. 

The relevant opacity for Equation \ref{equ-msigma} is the electron scattering opacity. For thermal motions significantly less than the rest mass energy, T$<<$ 10$^9$ K, this opacity is independent of photon frequency and is simply $\kappa$ = $\frac{n_e \sigma_e}{\rho}$, where $\sigma_e$ is the Thomson cross-section. It is unlikely that these terms have a significant dependence on $\sigma$ or on whether the galaxy is a central or a satellite. 

The static `protogalaxy' models used analytically implicitly assume that the black hole can accrete gas at an unencumbered rate which leads to no dependence of $\Gamma$ on $\sigma$ or \Mbh. Nonetheless, the Bondi-Hoyle accretion rate is $\propto$ M$^2_{\bullet}$ while the Eddington rate is $\propto$ \Mbh. Thus, if the black hole growth is accretion limited, a Bondi-Hoyle accreting black hole will have a $\Gamma$ variation with its mass. However, assuming this $\Gamma$ variation does not occur, the last possibility for a $\sigma$ dependence is the gas fraction, $f_g$.

\subsubsection{The dependence of gas fraction on $\sigma$}

Satellite galaxies infalling into groups and clusters experience ram pressure stripping of their hot gaseous halo and, perhaps, of the cold gas from the inner regions of the galaxy. These satellite galaxies are known to be deficient in gas and are not forming a significant amount of stars  \citep{HGC_84, Kenney_89, Boselli_97, Rasmussen_2012, Jaffe_2012, fabello_2012}. Thus, the gas fraction of satellite galaxies is likely universally low, with no significant dependence on $\sigma$. In contrast, central galaxies have declining gas fractions with increasing mass  \citep{Saintonge_coldgassI, catinella_2012}. This implies that the $f_g$ $\propto$ $\sigma^{-\alpha}$ where $\alpha$ depends on exactly how gas fractions decline with mass. Unfortunately, this is the wrong dependence to explain the observed scaling relations. However, the current f$_g$ - $\sigma$ relation is not necessarily the relevant quantity if black hole growth has been quiescent for some time. The f$_g$ at the time of the last major growth phase could be constant with velocity dispersion, and the current f$_g$ - $\sigma$ relation is a consequence of the resultant AGN outflows.

Recently, \citet{Zubovas_2012} explored the effect of $f_g$ changes due to gas depletion from star formation and gas replenishment in clusters. They suggested that the dominant form of gas loss in both dense and isolated environments is the conversion of gas to stars. They argued that the ram pressure stripping of galaxies in dense environments results only in the removal of gas on the outskirts of the galaxy, which lowers the star formation rate but does not significantly change the central BH mass required to drive an outflow. They further argued that cooling flows can replenish cold gas in some cluster galaxies (especially BCGs). This would actually increase the gas fractions in cluster galaxies and thus require higher BH masses at a given velocity dispersion. Thus, they expect an offset in the relation for cluster galaxies and isolated galaxies, but with the same slope.\footnote{\citet{Zubovas_2012} actually suggest the existence of 4 offset \Mbhrel\ relations: isolated spirals, cluster spirals, isolated ellipticals and cluster ellipticals which result from the combination of higher f$_g$ in clusters and larger bulge sizes (requiring longer quasar `on' phase) in ellipticals.} It is worth pointing out that we do see this offset when dividing the sample by halo mass, which may actually help to explain that relation. While this qualitative agreement is interesting, their requirement that cluster galaxies (which include satellite galaxies) have higher gas fractions than similar field galaxies (centrals) is hard to reconcile with the observations of gas in galaxies \citep{HGC_84, catinella2013}. It appears difficult to explain the different slopes in centrals and satellites by using secular growth as a guide.

\subsection{Merger driven growth of black holes}

Mergers are an important part of galaxy growth and a direct consequence of a \LCDM\ universe. For our purposes, the important quantity is the merger rate for centrals and satellites as a function of velocity dispersion (or some other mass indicator). Observationally, the merger rate is always difficult to measure because of its transient nature and the uncertain timescale of its disruption features. To add to this uncertainty, low mass satellite galaxies naturally have more close pairs than similar mass central galaxies and thus have substantial `projection effects'. The merger rate can be directly measured in the two environments through cosmological dark matter simulations which show that satellite-central mergers are more likely for non-dwarf satellite galaxies than are satellite-satellite mergers \citep{angulo09}. Assuming that baryons do not significantly alter this picture, then satellites live a relatively quiet life until they are subsumed by the central galaxy. However, given the uncertainty, it is worth exploring whether the nature of black hole growth in mergers differs between centrals and satellites.

There are two processes important for the black hole scaling relations in the context of mergers. Firstly, during a merger, the central black holes in each of the galaxies can also merge, or possibly be ejected from the dynamics. This will create differential satellite/central relations only if the frequency of each occurrence is different. Potentially the higher gas fractions in central galaxies could damp this effect, leading to more growth in centrals. Again, however, this would imply central galaxies have a shallower slope than satellite galaxies, in contradiction to our results. Secondly, if mergers induce bursts of star formation they could also drive gas towards the central black hole for continued growth \citep{sanders96, hopkins}. If this burst phase of star formation driven AGN growth is significantly different from the regular growth of black holes, then it could lead to a differential evolution of the central and satellite relations.

\subsection{The accelerated growth of black holes in low mass satellites}

If we assume that the \Mbhrel\ relation of centrals and satellites does not evolve with redshift, then low mass central galaxies must grow their black holes when they become satellite galaxies. The linear fits for the full BH sample give a satellite black hole mass of 2.0$\times$10$^7$ \Mdot\ for a $\sigma$ = 120 \kms\ galaxy while the corresponding black hole mass of central galaxies is slightly more than a third of that (7.3$\times$10$^6$ \Mdot). The situation switches at higher velocity dispersion, such that central galaxies of $\sigma$ = 320 \kms\ have black hole masses nearly twice those of satellite galaxies (3.8$\times$10$^9$\Mdot\ vs 2.1$\times$10$^9$\Mdot).

Given that low mass galaxies have considerable gas fractions, it may be that the difference between central and satellite relations is driven by these galaxies alone. A plausible scenario for these observed trends is that low mass (and therefore gas rich) galaxies have a phase of increased black hole growth during the transformation from a star forming central to a passive satellite galaxy. These passive satellites eventually merge with the central galaxy in the halo causing the relative increase in black hole mass at high velocity dispersion. It is important to remember that most of the merger accumulated mass occurs from minor mergers with objects 1/10 the size of the main galaxy \citep{lacey_cole}. These many minor mergers with over-massive satellite black holes are likely to lead to an over-massive central black hole in massive central galaxies. In contrast, high mass galaxies which become satellites do not have gas available for an additional growth of the black hole and thus have black hole masses similar to centrals.

We can obtain an estimate for how much of the mass of a 320 \kms central galaxy must have come from satellites of 1/10th the mass to explain its massive black hole without any accelerated growth within the galaxy. \citet{bernardi_paper2} showed that early type galaxies follow a relation such that L$_z$ $\propto$ R$^{1.58}$, where L$_z$ is the Sloan z' band luminosity and R is the effective radius of the galaxy. Assuming that these galaxies are in virial equilibrium and that the z band mass to light ratio is constant, then $\sigma$ $\propto$ L$_z^{0.61}$. As such, a galaxy which is 10 times more luminous has a $\sim$ 4 times higher velocity dispersion. As mentioned, a 320 \kms\ central galaxy has a black hole twice the value of a satellite black hole at the same velocity dispersion, while a satellite galaxy with 1/10 the luminosity, $\sigma$ = 75 \kms\ , would house a black hole 6 times larger than that of a central of the same velocity dispersion. Thus, if only 20$\%$ of the mass of a 320 \kms central galaxy has been made up by the accretion of satellites of 1/10th the mass, the mergers of their black holes will naturally produce a central black hole mass twice the size of a black hole in a satellite of the same mass. This amount of satellite accretion is easily accommodated by semi-analytic models \citep{delucia_bcg}. In effect, we require only that gas rich, low mass galaxies to have a phase of intense black hole growth to produce the relations we see.

\section{Conclusions}\label{sec-concl}
We have calculated four environmental indicators for a sample of galaxies with direct measurements of black hole mass, velocity dispersion, bulge luminosity and bulge mass. By examining how the scaling relations depend on these environmental parameters we arrive at the following conclusions. 

\begin{itemize}

\item The strongest dependence on environment of the \Mbh - $\sigma$ relation is seen when a division by satellite or central designation is made. While the vertical offsets of the relation of both samples are similar, central galaxies have a significantly steeper slope ( 6.38 $\pm$ 0.49 vs 4.91 $\pm$ 0.49).
\item The differential in central/satellite slopes is due not to an underlying difference in their varying fraction of early type and late type galaxies, since the slopes are nearly identical after the removal of spirals from both samples. The differential central/satellite slope remains even after removing the known Brightest Cluster Galaxies from the samples. 
\item The steeper central galaxy slope also occurs in the \Mbh\ - \Mbulge\ and \Mbh\ - \Mlum\ relations. The smaller sample sizes of these relations mean the significance of the trends is less in these relations. 
\item A simple model of secular black hole growth in which the black hole mass is coupled to the current gas fraction predicts that central galaxies would have shallower slopes than satellites, in contrast to our results. 
\item We suggest that gas rich, low mass galaxies experience a period of accelerated black hole growth during the transition from central to satellite. These galaxies, and their large black holes, eventually merge with massive central galaxies, leading to more massive black holes in massive central galaxies. 
\end{itemize}

The strong environmental dependence of black hole scaling relations provides a new test for models of the co-evolution of black holes and their galaxies. We suggest the most progress will be made by increasing the number of observations of low velocity dispersion, bulge mass and bulge luminosity isolated central galaxies.

\section*{Acknowledgments} 

We thank the anonymous referee for their detailed and helpful comments. We also thank Huub R\"{o}ttgering, George Miley, Alister Graham and Andrew King for useful comments.

\bibliography{../../ms}
\appendix
\section{Table of environmental parameters}
In table \ref{tab_params}, we present a full list of the environmental parameters derived and used in this study. 
\onecolumn
\begin{center}
\begin{longtable}{lcccc}
\hline
Galaxy & 4th Nearest Neighbour & Number density & Halo mass & Central/Satellite \\
& (Mpc) & (\# galaxies) & (Log$_{10}$(\Mdot)) & \\
\hline
 A1836 BCG &    3.48  &  18  & N/A    &        Central \\
 A3565 BCG &    2.03  &  34  &  13.46    &        Central \\
    IC1459 &    6.62  &   5  &  13.46    &        Central \\
  N224 (M31) &   $\infty$  &   2  &  11.33    &        Central \\
      N524 &    6.96  &   5  &  13.12    &        Central \\
      N821 &    5.26  &   7  & ---    &        Central \\
     N1023 &    9.89  &   4  &  13.20    &        Central \\
     N1194 &    8.19  &   8  & ---    &        Central \\
     N1300 &    2.88  &  16  & ---    &        Satellite \\
     N1316 &    1.33  &  15  &  13.95    &        Central \\
     N1332 &    4.89  &  14  &  13.96    &        Satellite \\
     N1374 &    0.33  &  15  &  13.95    &        Satellite \\
     N1399 &    0.86  &  14  &  13.95    &        Satellite \\
     N1407 &    4.39  &  12  &  13.96    &        Central \\
     N1550 &    3.91  &  16  &  14.02    &        Central \\
     N2273 &   10.57  &   2  & ---    &        Central \\
     N2549 &    8.82  &   4  & ---    &        Central \\
     N2787 &    6.99  &  13  & ---    &        Satellite \\
     N2960 &   10.58  &   3  & ---    &        Central \\
 N3031 (M81) &    8.48  &   5  &  12.30    &        Central \\
     N3091 &    6.24  &   5  & ---    &        Central \\
     N3115 &    6.27  &  22  & ---    &        Central \\
     N3227 &    8.51  &  12  &  11.97    &        Central \\
     N3245 &   11.26  &   1  &  11.99    &        Central \\
     N3368 &    4.53  &  27  &  13.29    &        Satellite \\
     N3377 &    4.74  &  19  &  13.29    &        Satellite \\
N3379 (M105) &    4.55  &  27  &  13.29    &        Satellite \\
     N3384 &    4.87  &  19  &  13.29    &        Satellite \\
     N3393 &    3.31  &  14  &  12.94    &        Central \\
     N3489 &    4.54  &  17  &  13.29    &        Satellite \\
     N3585 &    4.92  &   8  &  12.84    &        Central \\
     N3607 &    7.21  &  14  & 13.40     &       Central \\
     N3608 &    7.22  &  19  &  13.40    &        Satellite \\
     N3842 &    0.79  &  31  &  14.70    &        Central \\
     N3998 &    7.85  &   7  &  14.17    &        Satellite \\
     N4026 &    6.38  &  11  &  14.17    &        Satellite \\
     N4258 &    4.21  &  15  &  14.17    &        Satellite \\
     N4261 &    2.51  &  21  & ---    &        Central \\
     N4291 &    6.45  &   6  & ---    &        Satellite \\
     N4342 &    2.05  &  17  &  15.00    &        Satellite \\
 N4374 (M84) &    1.38  &  20  &  15.00    &        Satellite \\
     N4388 &    1.04  &  19  &  15.00    &        Satellite \\
     N4459 &    0.85  &  23  &  15.00    &        Satellite \\
 N4472 (M49) &    1.17  &  22  &  15.00    &        Central \\
     N4473 &    0.67  &   33 &   15.00   &         Satellite \\
 N4486 (M87) &    0.53  &  25 &   15.00    &        Satellite \\
    N4486A &    0.58  &  24 &   15.00    &        Satellite \\
     N4564 &    0.49  &  24 &   15.00    &        Satellite \\
N4594 (M104) &    1.96  &  36 &  ---    &        Central \\
     N4596 &    0.76  &  32 &  ---    &        Satellite \\
 N4649 (M60) &    1.17  &  23 &   15.00    &        Satellite \\
     N4697 &    1.35  &  38 &   13.63    &        Central \\
 N4736 (M94) &    2.54  &  16 &   14.17    &        Central \\
 N4826 (M64) &    1.42  &  17 &  ---    &        Central \\
     N4889 &    1.52  &  29 &   14.88    &        Central \\
     N5077 &    3.82  &   7 &  ---    &        Central \\
N5128 (CenA) &    2.24  &  28 &  ---    &        Central \\
     N5516 &    6.76  &   8 &   13.42    &        Central \\
     N5576 &   10.14  &   3 &   12.34    &        Satellite \\
     N5845 &   15.16  &   3 &   13.66    &        Satellite \\
     N6086 &    5.74  &  13 &  N/A    &        Central \\
     N6251 &    7.99  &   5 &  ---    &        Central \\
     N6264 &    0.54  &  32 &  N/A    &        Satellite \\
     N6323 &    1.17  &  26 &  ---    &        Central \\
     N7052 &   11.93  &   3 &  ---    &        Central \\
     N7582 &    5.41  &   7 &   12.14    &        Central \\
     N7619 &    4.33  &   9 &   13.94    &        Central \\
     N7768 &    3.27  &  16 &   12.76    &        Central \\
     U3789 &   12.74  &   1 &  ---    &        Satellite \\
 \hline

\caption{A compilation of the environmental parameters defined in \textsection \ref{sec-enviro}. Briefly, for each galaxy the 4th nearest neighbour distance, 10 Mpc number density, halo mass and whether the galaxy is a central or satellite galaxies is listed. Galaxies outside the volume of the group catalog have a halo mass of `N/A', while those within the volume but which are not a group member are listed as `---'.}
\label{tab_params}
\end{longtable}
\end{center}
\twocolumn

\end{document}